\documentclass[conference]{IEEEtran}
\IEEEoverridecommandlockouts
\usepackage{bm}
\usepackage{amsmath,amssymb,amsfonts}
\usepackage{graphicx}
\graphicspath{{sims/}}
\usepackage{textcomp}
\usepackage{color,xcolor}
\usepackage{cite}

\usepackage[numbers,sort&compress]{natbib}

\newcommand{\est}[1]{{\mathcal E}\!\left\{#1\right\}}

\newcommand{\rhant}{
\mbox{$ \; \circ\hspace*{-.5pt}\rule[2pt]{10pt}{.5pt}\!\bullet\; $}}
\newcommand{\lhant}{
\mbox{$\; \bullet\!\rule[2pt]{10pt}{.5pt}\hspace*{-.5pt}\circ \; $}}
\newcommand{\Real}[0]{\mathbb R}

\newcommand{\Complex}[0]{\mathbb C}

\newcommand{\ve}[1]{{\mathbf{#1}}}

\newcommand{\ive}[1]{{\boldsymbol{#1}}}

\newcommand{\Vektor}[1]{{\mathbf{#1}}}
\newcommand{\PVektor}[1]{{\boldsymbol{#1}}}
\newcommand{\Matrix}[1]{{\mathbf{#1}}}
\newcommand{\PMatrix}[1]{{\boldsymbol{#1}}}
\newcommand{\Herm}[0]{^{\mathrm{H}}}

\newcommand{\PH}[0]{^{\mathrm{P}}\!}
\newcommand{\Trans}[0]{^{\mathrm{T}}}

\newcommand{\diag}[1]{\mbox{diag}\!\left\{#1\right\}}
\newcommand{\trace}[1]{\mbox{tr}\!\left\{#1\right\}}

\newcommand{\ee}{\mathrm{e}}

\newcommand{\boldsigma}{\bm{\mathit{\Sigma}}}
\newcommand{\jj}{\mathrm{j}}
\newcommand{\ejo}{\ee^{\jj\Omega}}
\newcommand{\ejk}[1]{\ee^{\jj\Omega_{#1}}}

\newtheorem{example}{Example}

\DeclareSymbolFont{symbolsC}{U}{ntxsyc}{m}{n}
\SetSymbolFont{symbolsC}{bold}{U}{ntxsyc}{b}{n}
\DeclareMathSymbol{\multimapdotbothB}{\mathrel}{symbolsC}{24}

\newif\ifShowCorrections
\ShowCorrectionstrue
\ifShowCorrections
  \usepackage[normalem]{ulem}
  \definecolor{forgreen}{rgb}{0,0.6,0}
  \definecolor{orange}{RGB}{255,140,0}
  \definecolor{skyblue}{RGB}{100, 150, 235}
  
  \newcommand{\swc}[1]{{\color{forgreen}[#1]}}
\else

  \newcommand{\swc}[1]{}
\fi
\begin{document}

%%%%%%%%%%%%%%%%%%%%%%%%%%%%%%%%%%%%%%%%%%%%%%%%%%%%%%%%%%%%%%%%%
%
%  TITLE
%
%%%%%%%%%%%%%%%%%%%%%%%%%%%%%%%%%%%%%%%%%%%%%%%%%%%%%%%%%%%%%%%%%
\title{Impact of Estimation Errors of a Matrix of Transfer
  Functions onto Its Analytic Singular Values and Their Potential
  Algorithmic Extraction
  \thanks{The work of Mohammed Bakhit is funded by Mathworks and the
    University of Strathclyde. Faizan Khattak is the recipient of a
    scholarship of the Commonwealth Scholarship
    Commission.} }
\author{\IEEEauthorblockN{Mohammed A.~Bakhit, Faizan A.~Khattak, Ian
    K.~Proudler, and Stephan Weiss} \IEEEauthorblockA{Department of
    Electronic \& Electrical Engineering, University of Strathclyde,
    Glasgow G1 1XW, Scotland
    \\ \{mohammed.bakhit,faizan.khattak,ian.proudler,stephan.weiss\}@strath.ac.uk}}

\maketitle

%%%%%%%%%%%%%%%%%%%%%%%%%%%%%%%%%%%%%%%%%%%%%%%%%%%%%%%%%%%%%%%%%
%
%  ABSTRACT
%
%%%%%%%%%%%%%%%%%%%%%%%%%%%%%%%%%%%%%%%%%%%%%%%%%%%%%%%%%%%%%%%%%
\begin{abstract}
A matrix of analytic functions $\PMatrix{A}(z)$, such as the matrix of transfer functions in a multiple-input multiple-output (MIMO) system, generally admits an analytic singular value decomposition (SVD), where the singular values themselves are functions. When evaluated on the unit circle, for the sake of analyticity, these singular values must be permitted of become negative.  In this paper, we address how the estimation of such a matrix, causing a stochastic perturbation of $\PMatrix{A}(z)$, results in fundamental changes to the analytic singular values: for the perturbed system, we show that their analytic singular values lose any algebraic multiplicities and are strictly non-negative with probability one. We present examples and highlight the impact that this has on algorithmic solutions to extracting an analytic or approximate analytic SVD.
\end{abstract}

%%%%%%%%%%%%%%%%%%%%%%%%%%%%%%%%%%%%%%%%%%%%%%%%%%%%%%%%%%%%%%%%%
%
%  I. INTRODUCTION
%
%%%%%%%%%%%%%%%%%%%%%%%%%%%%%%%%%%%%%%%%%%%%%%%%%%%%%%%%%%%%%%%%%
\section{Introduction
  \label{sec:introduction}}

The singular value decomposition (SVD) is a well known and widely used
mathematical technique within the toolkit of linear algebra. It admits
a factorisation of any given matrix $\Matrix{A} \in \Complex^{M\times
  L}$ into a diagonalised form via unitary matrices~\cite{strang80a,
  golub96a}. In signal processing, this factorization has found
extensive applications summarised in e.g.~\cite{deprettere89a,
  vaccaro91a}. Specifically in signal processing for multiple-input
multiple-output communications, the SVD enables to determine precoding
and equalisation matrices that decouple a narrowband communications
channel, i.e.~a matrix comprising of complex gain factors, in order to
achieve system optimality in various senses~\cite{vu07a}. The desire
to extent this utility to the broadband case where the channel matrix
contains impulse responses --- or, in the $z$-domain, transfer
functions --- has given rise to the investigation and application of a
polynomial SVD~\cite{mcwhirter07a, ta07a, hanafy08a, zamiri08a,
  foster10a, moret11a, mestre14a, ahrens15a, perez16a, weiss23c,
  khattak23b, khattak23c, bakhit23a, khattak24a}.
  
In a polynomial matrix SVD~\cite{mcwhirter07a}, the SVD factors,
i.e.~the singular values and left- and right-singular vectors,
themselves become frequency-dependent. The left- and right-singular
vectors, which form so-called paraunitary matrices and implement
lossless filter banks~\cite{vaidyanathan93a}, can be approximated by
matrices of finite impulse response filters~\cite{foster06a, ta07d,
  corr15a, corr15b}. The order of these filters then contributes to
the implementation cost of any application; hence it is advantageous
to find polynomial SVD factorisations that are as compact in order as
possible.  SVD algorithms in~\cite{mcwhirter07a, foster10a,
  mcwhirter10b} are based on a set of algorithms that either favour or
even guarantee~\cite{mcwhirter16a} spectrally majorised singular
values; these converge towards piecewise analytic functions and are
therefore difficult to approximate. More recently, the existence of an
analytic SVD has been established~\cite{weiss24b, barbarino23a}, which
postulates significantly smoother and therefore lower order SVD
factors than may be achievable with current techniques. The extraction
of analytic SVD factors therefore seems attractive, with initial
attempts reported in~\cite{khattak23b, khattak23c, bakhit23a,
  khattak24a} which operate analogously to the principles behind
analytic EVD algorithms in~\cite{tohidian13a, weiss19a, weiss20c,
  weiss21a, weiss23b, khattak23a, khattak24b, khattak24d}.

In practice, for example in a communications system where the channel
needs to be estimated from a finite amount of data, the system to
which we want to apply an analytic SVD will be perturbed by a random
component.  In related work for the analytic eigenvalue decomposition
of a space-time covariance matrix~\cite{khattak22c}, a profound
challenge has occurred: even if the space-time covariance matrix is
subjected to only very small estimation errors~\cite{khattak22a},
their perturbation effect on the analytic eigenvalues cannot be
neglected. According to~\cite{khattak22a}, unless the transition to an
infinite sample size and a zero estimation error is made, the
eigenvalues of the perturbed matrix will be spectrally majorised even
if the unperturbed eigenvalues are not.  Missing the smoothness of the
ground truth solution means that any identified solution will require
a significantly higher approximation order for the EVD factors,
resulting in potentially high implementation costs.

Therefore, in this paper we want to investigate how approximation
errors in a matrix of analytic functions will impact on its analytic
SVD, and specifically its analytic singular values.  Below, in
Sec.~\ref{sec_ana_svd} we review the analytic SVD, followed by a
discussion on how the estimation perturbs a system matrix in
Sec.~\ref{sec:A_Perturb}.  Its impact of the perturbation of the
singular values is laid out in Sec.~\ref{sec:Sigma_perturb}, which
contains the main findings of this paper: the singular values of a
random perturbed matrix are non-negative and spectrally majorised with
probability one. For this, we provide some examples in
Sec.~\ref{sec:sims} and discuss consequences Sec.~\ref{sec:concl}.

%%%%%%%%%%%%%%%%%%%%%%%%%%%%%%%%%%%%%%%%%%%%%%%%%%%%%%%%%%%%%%%%%
%
%  II. ANALYTIC SVD
%
%%%%%%%%%%%%%%%%%%%%%%%%%%%%%%%%%%%%%%%%%%%%%%%%%%%%%%%%%%%%%%%%%
\section{Analytic Singular Value Decomposition 
  \label{sec_ana_svd}}

%%%%%%%%%%%%%%%%%%%%%%%%%%%%%%%%%%%%%%%%%%%%%%%%%%%%%%%%%%%%%%%%%
%
%%%%%%%%%%%%%%%%%%%%%%%%%%%%%%%%%%%%%%%%%%%%%%%%%%%%%%%%%%%%%%%%%
\subsection{Standard Singular Value Decomposition
   \label{sec:svd} }

For any given matrix $\Matrix{A}\in\Complex^{M\times L}$, without loss
of generality\footnote{Otherwise we address $\Matrix{A}\Herm$ instead
of $\Matrix{A}$.} assuming $M\leq L$, there is a singular value
decomposition \citep{golub96a} such that
\begin{align}
\label{svd_decomp}
\Matrix{A}=\Matrix{U}\PMatrix{\Sigma}\Matrix{V}\Herm \;,
\end{align}
where $\ve{U}$ $\in\Complex^{M\times M}$ and $\ve{V}$
$\in\Complex^{L\times L}$ are the left- and right-singular vectors.
The quantity $\PMatrix{\Sigma}$ $\in\Real^{M\times L}$ in
\eqref{svd_decomp} is a diagonal matrix containing the 
unique singular values $\sigma_m$, $m=1,\dotsc,M$, such that
\begin{align}
\label{eq_spec_maj}
 \sigma_m \geq \sigma_{m+1}\geq 0, \qquad \forall m=1,\dotsc,(M-1) \; .
\end{align}
As per the standard definition of the SVD, the singular values are
constrained to be non-negative. If there are singular values $\sigma_m
= \sigma_{m+1} = \cdots = \sigma_{m+C-1}$, we speak of these singular
values as possessing a $C$-fold algebraic multiplicity~\cite{golub96a,
  kato80a}.

%%%%%%%%%%%%%%%%%%%%%%%%%%%%%%%%%%%%%%%%%%%%%%%%%%%%%%%%%%%%%%%%%
%
%%%%%%%%%%%%%%%%%%%%%%%%%%%%%%%%%%%%%%%%%%%%%%%%%%%%%%%%%%%%%%%%%
\subsection{Extension to Matrices of Analytic Functions}

Extending \eqref{svd_decomp} to polynomial matrices, or generally to
matrices of functions $\PMatrix{A}(z): \Complex \rightarrow
\Complex^{M\times L}$ that are analytic in $z$, requires an analytic
SVD~\cite{barbarino23a,weiss24b}. Unless the system $\PMatrix{A}(z)$
contains any multiplexing operation, such as implicit block
filtering~\cite{scheuermann81a, vaidyanathan90a, akansu98a}, the
analytic SVD takes the form
\begin{align}
\label{eq_ana_svd}
\ive{A}(z)=\ive{U}(z)\boldsigma(z)\ive{V}\PH(z) \; .
\end{align}
In \eqref{eq_ana_svd}, $\ive{U}(z)\in\mathbb{C}^{M\times M}$ and
$\ive{V}(z)\in\mathbb{C}^{L\times L}$ are the left- and right-singular
vectors, respectively. Note that $\ive{U}(z)$ and $\ive{V}(z)$ are
paraunitary matrices i.e.~$\ive{U}(z)$
$\ive{U}(z)\PH=\ve{I},~\ive{V}(z)\ive{V}\PH(z)=\ve{I}$, whereby the
parahermitian operator $\{\cdot\}\PH$ implies a Hermitian
transposition and time reversal such that $\ive{U}\PH(z) =
\{\ive{U}\Herm(1/z^{\ast})\}\Herm$.  The diagonal matrix
$\boldsigma(z) = \diag{\sigma_1(z), \dotsc, \sigma_M(z)}
\in\mathbb{C}^{M\times L}$ contains the singular values $\sigma_m(z)$,
$m=1,\dotsc,M$. In order to admit analyticity of the SVD factors in
\eqref{eq_ana_svd}, the singular values must be permitted to become
negative on the unit circle, i.e.~the restriction of non-negativity as
known from the standard SVD in Sec.~\ref{sec:svd} must be
dropped. This is known for the case where a matrix $\PMatrix{A}(t)$ is
a function in a real parameter $t\in \Real$~\cite{demoor89a,
  bunsegerstner91a} as well as for the case of a dependency of
$\PMatrix{A}(z)$ on a complex-valued variable
$z\in\Complex$~\cite{barbarino23a, weiss24b}.

\begin{example} \label{ex:ex1}
  Consider the matrix
  \begin{align}
    \PMatrix{A}(z) & = \frac12 \left[ \begin{array}{c}
        (\tfrac14-\jj)z + 1 + (\tfrac14+\jj)z^{-1} \\
        (\tfrac14+\jj)z + 1 + (\tfrac14-\jj)z^{-1}
       \end{array} \right. \nonumber \\     
     &  \qquad \qquad \qquad \left. \begin{array}{c}
                   -(\tfrac14+\jj)z - 1 - (\tfrac14-\jj)z^{-1} \\
                   -(\tfrac14-\jj)z - 1 - (\tfrac14+\jj)z^{-1}
      \end{array} \right]  \; ,
    \label{eqn:A_ex1}
  \end{align}
  which has the analytic singular values $\sigma_1(z) = \tfrac14z + 1
  + \tfrac14z^{-1}$ and $\sigma_2(z) = -\jj z + \jj z^{-1}$. One
  option for the analytic left- and right-singular vectors is
  $\PMatrix{U}(z) = [1, \; 1; \; 1 \; -1]/\sqrt{2}$ and
  $\PMatrix{V}(z) = z^{-1}[1,\; 1; -1, \; 1]/\sqrt{2}$. The evaluation
  of the analytic singular values on the unit circle,
  $\sigma_m(\ejo)$, $m=1,2$, is shown in Fig.~\ref{fig:ex1}.  \hfill
  $\triangle$
  \begin{figure}
    \includegraphics[trim=0 0 0 0,clip,width=\columnwidth]{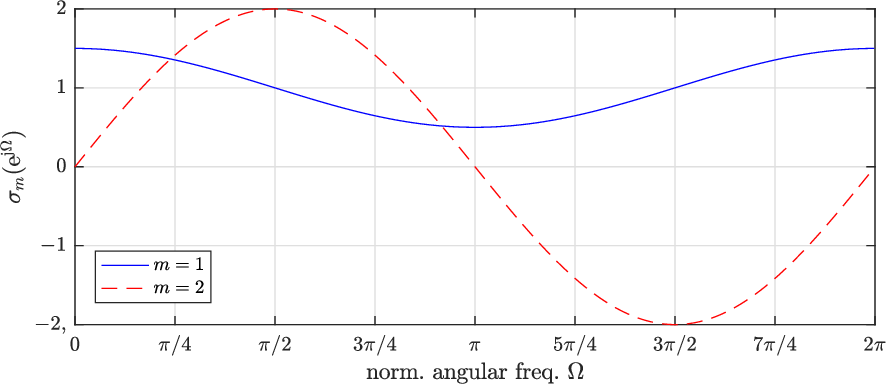}
    \caption{Example of analytic singular values of matrix $\PMatrix{A}(z)$ in
      \eqref{eqn:A_ex1}.
      \label{fig:ex1}}
  \end{figure}  
\end{example}

Analyticity is an important property, since analytic functions can be
approximated arbitrarily closely by shift and truncation
operations~\cite{weiss23b}. If analyticity is denied, the singular
values may only form piece-wise analytic functions, whereby
non-differentibilities arise which are difficult to approximate even
by polynomial factors of very high order~\cite{weiss21a}; this has
consequences for the implementation of such factors for e.g.~precoding
and equalisation of MIMO communications
channels~\cite{ta07a, moret11a}, where high approximation orders for
$\PMatrix{U}(z)$ and $\PMatrix{V}(z)$ result in their realisation via
filter banks~\cite{vaidyanathan93a} requiring very long filters.  .

%%%%%%%%%%%%%%%%%%%%%%%%%%%%%%%%%%%%%%%%%%%%%%%%%%%%%%%%%%%%%%%%%
%
%%%%%%%%%%%%%%%%%%%%%%%%%%%%%%%%%%%%%%%%%%%%%%%%%%%%%%%%%%%%%%%%%
\subsection{Ambiguities of the Analytic SVD}

To explore the ambiguities of the analytic SVD in \eqref{eq_ana_svd},
consider its equivalent representation using a sum of orthogonal
rank-one terms,
\begin{align}
    \PMatrix{A}(z) = \sum_{m=1}^{M} \Vektor{u}_m(z) \sigma_m(z)
    \Vektor{v}_m\PH(z) \; .
\end{align}  
With the analytic SVD maintaining $\sigma_m(\ejo) \in \Real$ but
dropping $\sigma_m(\ejo)\geq 0$, if $\sigma_m(z)$ is a valid $m$th
analytic singular vector, then so is $-\sigma_m(z)$.  We can therefore
obtain a different but equally valid $m$th analytic singular value
$\sigma^\prime_m(z)$ and corresponding analytic left- and
right-singular vectors $\PVektor{u}^\prime_m(z)$ and as
$\PVektor{v}^\prime_m(z) $ as
\begin{align}
  \sigma^\prime_m(z) & = \ee^{\jj\pi\zeta_m} \sigma_m(z)
      \label{eqn:ambiguity1} \\
  \PVektor{u}^\prime_m(z) & = \varphi_m(z) \PVektor{u}_m(z) \\
  \PVektor{v}^\prime_m(z) & = \varphi_m(z) \ee^{\jj\pi\zeta_m} \PVektor{v}_m(z) \; ,
      \label{eqn:ambiguity3}
\end{align}  
where $\zeta_m\in\{0,1\}$ and an allpass $\varphi_m(z)$ are
arbitrary. If we exclude the case of identical singular values, such
that $\sigma_m(\ejo) = \sigma_\mu(\ejo) \forall \Omega$ for some $\mu
\neq m$, where further choices arise~\cite{weiss24b},
\eqref{eqn:ambiguity1} -- \eqref{eqn:ambiguity3} expresses the
ambiguity of the analytic SVD.

%%%%%%%%%%%%%%%%%%%%%%%%%%%%%%%%%%%%%%%%%%%%%%%%%%%%%%%%%%%%%%%%%
%
%  III.  RANDOM PERTURBATION OF ANALYTIC MATRIX
%
%%%%%%%%%%%%%%%%%%%%%%%%%%%%%%%%%%%%%%%%%%%%%%%%%%%%%%%%%%%%%%%%%
\section{Random Perturbation of An Analytic Matrix 
  \label{sec:A_Perturb}}

In this section, we briefly highlight how a matrix $\Matrix{A}(z)$ may
be randomly perturbed e.g.~through the process of estimation. We
provide an example, where $\Matrix{A}(z)$ is obtained by system
identification via the Wiener solution. This estimate
$\hat{\PMatrix{A}}(z)$ will appear randomly perturbed with respect to
the $\PMatrix{A}(z)$, and we state factors that influence the variance
of this estimate in case of a system identification process involving
Gaussian distributed signals.

%%%%%%%%%%%%%%%%%%%%%%%%%%%%%%%%%%%%%%%%%%%%%%%%%%%%%%%%%%%%%%%%%
%  III.A  System Identification Setup
%%%%%%%%%%%%%%%%%%%%%%%%%%%%%%%%%%%%%%%%%%%%%%%%%%%%%%%%%%%%%%%%%
\subsection{System Identification Setup}

\begin{figure}[t]
  \centerline{\includegraphics[width=0.8\columnwidth]
        {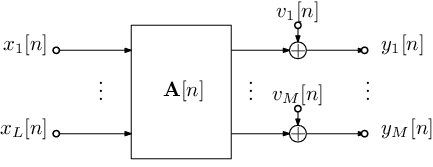}}
  \caption{System matrix $\Matrix{A}[n] \rhant \PMatrix{A}(z)$ excited
    by inputs $x_{\ell}[n]$, $\ell=1,\dotsc,L$ source contributions and
    providing $M$ outputs $y_m[n]$ corrupted by additive noise
    $v_m[n]$, $m=1,\dotsc,M$.
    \label{fig:sys_id}}
\end{figure}

Instead of having direct access to a matrix of transfer functions
$\PMatrix{A}(z)$ with $L$ inputs and $M$ outputs, consider the
scenario where we can only measure the input-output behaviour of
$\PMatrix{A}(z)$ as illustrated in Fig.~\ref{fig:sys_id}.
Assume that we have $L$ zero-mean unit-variance uncorrelated sources
$x_{\ell}$, $\ell = 1,\dotsc,L$, contribute to the measurements at $M$
sensors via a matrix $\Matrix{A}[n] \in \Complex^{M\times L}$ of
impulse responses. For $\Matrix{A}[n]$, the element $a_{m,\ell}[n]$ in
the $m$th row and $\ell$th column,
\begin{align*}
\Matrix{A}[n]=  \left[\begin{matrix}
 {a}_{1,1}[n] & \cdots  & {a}_{1,L}[n] \\
    \vdots & \ddots & \vdots \\
    {a}_{M,1}[n & \cdots  & {a}_{M,L}[n]
\end{matrix}\right]
\end{align*}
is the impulse response connecting the $\ell$th source to the ${m}$th
sensor.  Thus the contribution of the ${m}$th sensor from all $L$
sources can be written as
\begin{align}
\label{contrubition_at_mth_sensor}
   y_{m}[n]=\sum_{\ell=1}^{L} a_{m,\ell}[n] \ast x_{\ell}[n]+\ve{v}_{m}[n] \; ,
\end{align}
where $\ast$ denotes the convolution operator and $v_{m}[n]$ is
additive spatially and temporally uncorrelated noise of variance
$\sigma^2_v$. Overall, using vectors $\Vektor{x}[n] = [x_1[n], \dotsc,
  x_L[n]]\Trans$, $\Vektor{v}[n] = [v_1[n],\dotsc,v_M[n]]\Trans$, and
$\Vektor{y}[n]=[y_1[n],\dotsc, y_M[n]]\Trans$, we can write
 \begin{align}
\label{vector_x}
\ve{y}[n]=\Matrix{A}[n] \ast \ve{x}[n]+\ve{v}[n] \; .
 \end{align}
 The aim is now to identify $\Matrix{A}[n]$ from a finite number of
 measurements $\Vektor{x}[n]$ and $\Vektor{y}[n]$, $n=0,\dotsc,(N-1)$.
 
%%%%%%%%%%%%%%%%%%%%%%%%%%%%%%%%%%%%%%%%%%%%%%%%%%%%%%%%%%%%%%%%%
%  III.B  Wiener Solution
%%%%%%%%%%%%%%%%%%%%%%%%%%%%%%%%%%%%%%%%%%%%%%%%%%%%%%%%%%%%%%%%%
 \subsection{Wiener Solution
    \label{sec:wiener}}

For general adaptive system identification system of $\Matrix{A}[n]$,
we adjust a matrix $\hat{\Matrix{A}}[n]$, excited by the same input
$\Vektor{x}[n]$ and generating a separate output $\hat{\Vektor{y}}[n]$. The aim is to adjust
$\hat{\Matrix{A}}[n]$ such that the mean square error 
\begin{align}
  \xi_{\mathrm{MSE}} & = \est{ \| \hat{\Vektor{y}}[n] - \Vektor{y}[n]
    \|^2_2} \nonumber \\
  & = \est{\| \hat{\Matrix{A}}[n] \ast \Vektor{x}[n] -
    \Vektor{y}[n] \|^2_2}
   \label{eqn:MSE}
\end{align}  
is minimised, where $\|\cdot\|_2$ is the $\ell_2$-norm and
$\est{\cdot}$ the expectation operator. Thus, we solve the problem
\begin{align}
  \hat{\Matrix{A}}_\ast[n] & = \mathrm{arg}\min_{\hat{\Matrix{A}}[n]}
     \xi_{\mathrm{MSE}} \; ,
\end{align}  
where $\hat{\Matrix{A}}_{\ast}[n]$ is the minimum mean square error
or Wiener solution~\cite{widrow85a, haykin91a}.
  
The Wiener solution involves a data covariance matrix derived from
$\Vektor{x}[n]$, and the cross-correlation between $\Vektor{x}[n]$ and
$\Vektor{y}[n]$. These correlation terms themselves must be estimated
from a finite number $N$ of samples of $\Vektor{x}[n]$ and
$\Vektor{y}[n]$, $0 \leq n < N$. For an unbiased estimator and
Gaussian data, the variance of these estimates depends on both the size
$N$ and the true correlation of the signals~\cite{delaosa19a}.
  
%%%%%%%%%%%%%%%%%%%%%%%%%%%%%%%%%%%%%%%%%%%%%%%%%%%%%%%%%%%%%%%%%
%  III.C  Variance of Estimation Error
%%%%%%%%%%%%%%%%%%%%%%%%%%%%%%%%%%%%%%%%%%%%%%%%%%%%%%%%%%%%%%%%%
\subsection{Variance of Estimation Error
   \label{sec:var}}

For the mean square error, note that inserting \eqref{vector_x} into
\eqref{eqn:MSE} yields
\begin{align}
  \xi_{\mathrm{MSE}} & = \est{ \| (\hat{\Matrix{A}}[n] - \Matrix{A}[n])
    \ast \Vektor{x}[n]
    + \Vektor{v}[n] \|^2_2}  \; .
  \label{eqn:xi2}
\end{align}     
The identification error
\begin{align}
  \Matrix{E}[n] & = \hat{\Matrix{A}}[n] - \Matrix{A}[n]
  \label{eqn:E1}
\end{align}
is the perturbation of $\Matrix{A}[n]$ in which we are interested.
Further, if $\Matrix{E}[n]$ has a temporal order $J$, we have
\begin{align}
  \Matrix{E}[n] \ast \Vektor{x}[n] = \left[\Matrix{E}[0], \dotsc,
    \Matrix{E}[J] \right] \cdot
  \left[ \begin{array}{c}
      \Vektor{x}[n] \\ \vdots \\ \Vektor{x}[n-J]
      \end{array} \right]  = \Matrix{E} \Vektor{x}_n
\end{align}
Assuming that the input signal, the noise, and with a wider stretch
the identification error $\Matrix{E}[n]$ are mutually independent,
\eqref{eqn:xi2} can be written as
\begin{align}
  \xi_{\mathrm{MSE}} & = \est{ \trace{ \Matrix{E} \Vektor{x}_n \Vektor{x}_n\Herm
      \Matrix{E}\Herm}} + \est{\trace{\Vektor{v}[n] \Vektor{v}\Herm[n]}} \nonumber \\
  & = \trace{\est{\Matrix{E}\Herm \Matrix{E}} \est{ \Vektor{x}_n \Vektor{x}\Herm_n}} +   \trace{\est{\Vektor{v}[n] \Vektor{v}\Herm[n]}} \; , \nonumber
\end{align}  
with $\trace{\cdot}$ the trace operator and exploiting
$\trace{\Matrix{ABC}} = \trace{\Matrix{CAB}}$. With $\est{\Vektor{x}_n
  \Vektor{x}_n} = \Matrix{I}_{JM}$ and $\est{\Vektor{v}[n]
  \Vektor{v}\Herm[n]} = \sigma^2_v \Matrix{I}_M$, the MSE 
simplifies to
\begin{align}
  \xi_{\mathrm{MSE}} & = \est{ \trace{\Matrix{E}\Herm \Matrix{E}}} + M\sigma^2_v 
     \nonumber \\  
     & = \est{\sum_n \| \Matrix{E}[n] \|^2_{\mathrm{F}}} + M\sigma^2_v  \; .
  \label{eqn:xi3}   
\end{align}
In \eqref{eqn:xi3}, the MSE consists of the minimum mean square error
components $M\sigma^2_v$ assuming that the estimate
$\hat{\Matrix{A}}[n]$ is not curtailed in length to incur truncation
errors. Any estimation errors due to a finite sample size $N$ as
discussed in Sec.~\ref{sec:wiener} will therefore contribute to an
offset of the MSE with respect to the MMSE by the variance of the
estimation error of $\Matrix{A}[n]$, i.e.~by $\est{\sum_n \|
  \Matrix{E}[n] \|^2_{\mathrm{F}}}$. In the following, we assume that
$\Matrix{E}[n]$ itself is random with its elements being Gaussian
distributed.

%%%%%%%%%%%%%%%%%%%%%%%%%%%%%%%%%%%%%%%%%%%%%%%%%%%%%%%%%%%%%%%%%
%
%  IV. Perturbation of Singular Values
%
%%%%%%%%%%%%%%%%%%%%%%%%%%%%%%%%%%%%%%%%%%%%%%%%%%%%%%%%%%%%%%%%%
\section{Perturbation of Singular Values
  \label{sec:Sigma_perturb}}

We are now interested in how a random perturbation of $\PMatrix{A}(z)
= \sum_n \Matrix{A}[n] z^{-n}$, or short $\PMatrix{A}(z) \lhant
\Matrix{A}[n]$, by a Gaussian term $\PMatrix{E}(z) \lhant
\Matrix{E}[n]$ impacts on the analytic singular values of the
perturbed matrix. We first assess the perturbation in isolated
frequency bins, before expanding our reasoning to the entire frequency
axis.

%%%%%%%%%%%%%%%%%%%%%%%%%%%%%%%%%%%%%%%%%%%%%%%%%%%%%%%%%%%%%%%%%
%  IV.A   Bin-Wise Perturbation
%%%%%%%%%%%%%%%%%%%%%%%%%%%%%%%%%%%%%%%%%%%%%%%%%%%%%%%%%%%%%%%%%
\subsection{Bin-Wise Perturbation
  \label{sec:bin_perturb}}

Based on the perturbation of $\Matrix{A}[n]$ in \eqref{eqn:E1}, we now
define the $z$-domain equivalent as
\begin{align}
  \hat{\PMatrix{A}}(z) & = \PMatrix{A}(z) + \PMatrix{E}(z) \; ,
\end{align}
where $\PMatrix{E}(z) \lhant \Matrix{E}[n]$ is the estimation error
whose variance has been assessed in Sec.~\ref{sec:var}.  With a
bin-wise view, the singular values $\varsigma_m$ of
$\hat{\PMatrix{A}}(z)|_{z=\ee^{\jj\Omega_0}}$ evaluated at a
normalised angular frequency $\Omega_0$, i.e.~via an SVD of
$\hat{\PMatrix{A}}(\ejk{0})$, are now stochastic quantities that obey
some probability distribution.  Therefore, in the case that the
singular values $|\sigma_m(\ejk{0})|$ of $\PMatrix{A}(\ejk{0})$ have a
$C$-fold algebraic multiplicity, $C$ singular values $\varsigma_\mu$,
$\mu=m,\dotsc,m+C-1$, are sampled from this distribution. As a result,
the singular values $\varsigma_m$ will be distinct with probability
one. This is analogous to the case of eigenvalues of a randomly
perturbed parahermitian matrices~\cite{khattak22c}. Thus, the
perturbed bin-wise singular values will, with probability one, not
possess any non-trivial algebraic multiplicities.

For small singular values, G.W.~Stewart in~\cite{stewart90b} states
that they ``tend to increase under perturbation, and the increment is
proportional to $\sqrt{M}$''. If $|\sigma_m(\ejk{0})|$ is the modulus
of a small singular value of $\PMatrix{A}(\ejk{0})$, then the square
of the corresponding singular value of $\hat{\PMatrix{A}}(\ejk{0})$
can be expanded as
\begin{align}
  \varsigma^2_m = (|\sigma_m(\ejk{0})| + \gamma)^2_m + \eta_m^2 \; ,
\end{align}  
where the terms $\gamma$ and $\eta$ are bounded
by~\cite{stewart79a,stewart84a} such that
\begin{align}
  |\gamma_m| & \leq \| \Matrix{P} \PMatrix{E}(\ejk{0}) \|_2
  \\ \inf{}_{\!2} \{\Matrix{P}_{\perp} \PMatrix{E}(\ejk{0})\} & \leq
  \eta_m \leq \| \Matrix{P}_\perp \PMatrix{E}(\ejk{0}) \|_2 \; .
  \label{eqn:eta}
\end{align}
The matrix $\PMatrix{P}$ is the orthogonal projection into the column
space of $\PMatrix{A}(\ejk{0})$, $\PMatrix{P}_\perp$ its complement,
and $\inf_2\{\Matrix{X}\}$ the smallest singular value of
$\Matrix{X}$~\cite{stewart90b}.

While particularly $\eta^2$ creates a bias for small
$|\sigma_m(\ejk{0})|$, for $\sigma_m(\ejk{0})=0$, the lower bound in
\eqref{eqn:eta} will be zero, and theoretically $\varsigma_m=0$ is
possible. Note that $\varsigma^2_m$ is also an eigenvalue of the
Hermitian matrix $\Matrix{R} = \PMatrix{A}(\ejk{0})
\PMatrix{A}\Herm(\ejk{0})$. In~\cite{azais04a,ratnarajah04a,anderson10a}
the distributions of small eigenvalues and the limits for the
condition number of a $\Matrix{R}$ for a Gaussian
$\PMatrix{A}(\ejk{0})$ are derived. Since the condition number of
$\Matrix{R}$ in e.g.~\cite{azais04a} has an upper bound, we must have
$\lambda_{\min}(\Matrix{R}) >0$, and therefore by implication
$|\varsigma_m|>0$.  Thus, a randomly perturbed matrix will, with
probability one, possess no zero singular values.

The probability distribution of singular values can be difficult to
express, see e.g.~\cite{shen01a}, but we provide two simple
examples.
\begin{example} Consider the degenerate case of $\Matrix{A}, \;
\Matrix{E} \in \Complex^{1 \times 1}$ but $\Matrix{A}=0$, and
$\Matrix{E}$ complex Gaussian distributed. Then we have $\sigma=0$ but
$\varsigma=|\Matrix{E}|$.  Thus, note that for any finite perturbation
$\Matrix{E}$, indeed $\varsigma>0$.  Further, given the complex Gaussian
distribution of $\Matrix{E}$, $\varsigma$ will be Rician
distributed. \hfill $\triangle$
\end{example}
\begin{example} Recall the matrix $\PMatrix{A}(z)$ from 
  Example~\ref{ex:ex1}. With a bin-wise evaluation at $\Omega_0=\pi$,
  Fig.~\ref{fig:hist} shows the histogram of the two singular values
  over $10^4$ different random perturbations . While
  $\sigma_2(\ee^{\jj\Omega_0})=0$, it is evident from
  Fig.~\ref{fig:ex1}(a) that the distribution of $\varsigma_2$ does
  not include zero.  The histograms in Fig.~\ref{fig:hist} also
  include Rician fits, which suggest that $\varsigma_m$, $m=1,2$, even
  for $\varsigma_m \gg 0$ still adhere to this distribution in good
  approximation. \hfill $\triangle$
\begin{figure}
  \parbox{0.43\columnwidth}{
    \includegraphics[trim=0 0 0 0,clip,width=0.43\columnwidth]{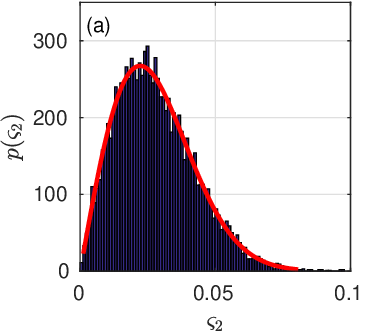}
  }\hfill    
  \parbox{0.5475\columnwidth}{
    \includegraphics[trim=0 0 0 0,clip,width=0.5475\columnwidth]{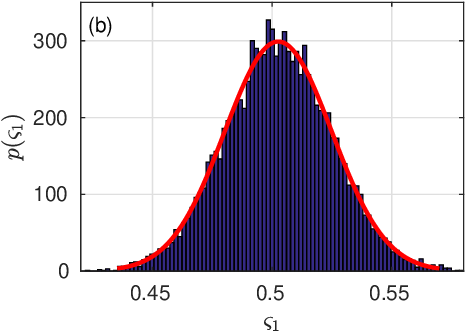}
  } 
  \caption{Histograms of perturbed zero singular values and fits by a
    Rician distribution (in red) for (a) $\varsigma_2$ and (b) $\varsigma_1$.
    \label{fig:hist}}
\end{figure}
\end{example}

Thus, overall the perturbed singular values will, with probability
one, have only trivial algebraic multiplicities and satisfy
$\varsigma>0$.

%%%%%%%%%%%%%%%%%%%%%%%%%%%%%%%%%%%%%%%%%%%%%%%%%%%%%%%%%%%%%%%%%
%  IV.B   
%%%%%%%%%%%%%%%%%%%%%%%%%%%%%%%%%%%%%%%%%%%%%%%%%%%%%%%%%%%%%%%%%
\subsection{Perturbed Analytic Singular Values}

In order to assess the singular values of the perturbed system
$\hat{\Matrix{A}}(z)$, note that $\hat{\Matrix{A}}(z)$ remains
analytic in $z\in\Complex$ and therefore admits an analytic singular
value decomposition~\cite{weiss24b}. Assessing these analytic singular
values $\hat{\sigma}_m(z)$ on the unit circle, at every frequency the
result from Sec.~\ref{sec:bin_perturb} has to hold.
As a consequence, the analytic singular values $\hat{\sigma}_m(\ejo)$
of $\hat{\PMatrix{A}}(z)$ will
\begin{enumerate}
  \item neither intersect each other
  \item nor cross the zero line
\end{enumerate}
with probability one.

As an interpretation, on the unit circle the analytic singular values
$\hat{\sigma}_m(\ejo)$ will approximate functions that arise from
extracting spectrally majorised versions of $\pm \sigma_m(\ejo)$.
Where they intersect --- and this includes the intersections of $\sigma_m(\ejo)$
with $-\sigma_m(\ejo)$ ---, a permutation occurs. We illustrate this via
the following simple example.

\begin{example} \label{ex:ex4}
  The system $\PMatrix{A}(z)$ from Example~\ref{ex:ex1} is now
  perturbed by a random term $\PMatrix{E}(z) \lhant \Matrix{E}[n]$ of
  the same order as $\PMatrix{A}(z)$. The perturbation is such that
  for each element $e_{m,\mu}[n]$, $m,\mu=1,2$ and $0\leq n \leq 2$,
  of $\Matrix{E}[n]$, $e_{m,\mu}[n] \sim \mathcal{CN}(0,\sigma^2_e)$
  with $\sigma^2_e=10^{-4}$. The singular values of
  $\hat{\PMatrix{A}}(z) = \PMatrix{A}(z) + \PMatrix{E}(z)$ are
  illustrated in Fig.~\ref{fig:ex5}, where the analytic singular
  values $\hat{\sigma}_m(\ejo)$ are linearly interpolated from a very
  high resolution bin-wise SVD evaluation. Fig.~\ref{fig:ex5}(b)
  highlights the loss of an algebraic multiplicity of
  $\sigma_m(\ejo)$, while the previous zero-crossing of
  $\sigma_2(\ejo)$ at $\Omega=\pi$ no longer occurs for
  $\hat{\sigma}_2(\ejo)$. \hfill $\triangle$
  \begin{figure}
  \includegraphics[trim=0 0 0 0,clip,width=\columnwidth]{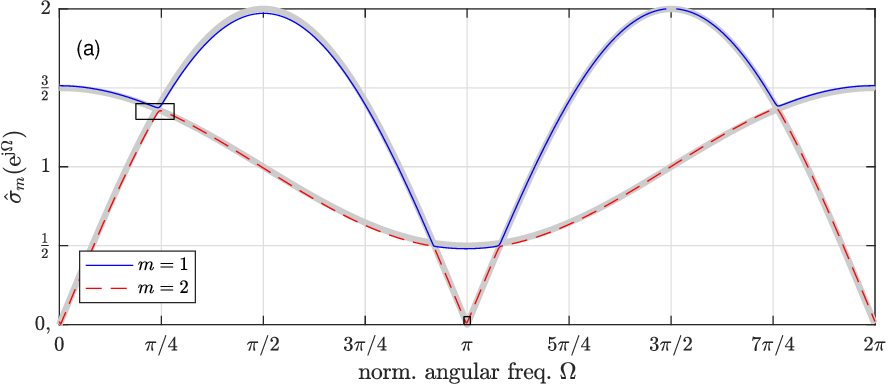}
  \includegraphics[trim=0 0 0 0,clip,width=\columnwidth]{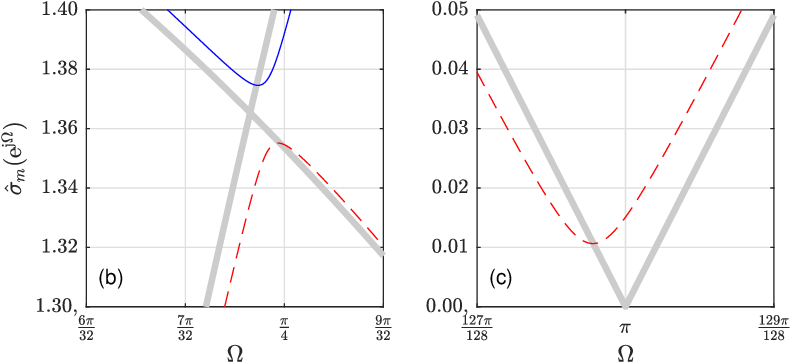}
  \caption{(a) analytic singular values $\hat{\sigma}_m(\ejo)$ of the
    randomly perturbed matrix $\hat{\Matrix{A}}(z)$, with the moduli
    of the ground truth analytic singular values $|\sigma_m(\ejo)|$
    underlaid in grey; zoomed version of this graph near (b) an
    algebraic multiplicity and (c) a zero crossing of the ground truth
    singular values.
    \label{fig:ex5}}
\end{figure}
\end{example}

%%%%%%%%%%%%%%%%%%%%%%%%%%%%%%%%%%%%%%%%%%%%%%%%%%%%%%%%%%%%%%%%%
%
%  V. EXAMPLES AND SIMULATIONS
%
%%%%%%%%%%%%%%%%%%%%%%%%%%%%%%%%%%%%%%%%%%%%%%%%%%%%%%%%%%%%%%%%%
\section{Examples and Simulation
  \label{sec:sims}}

%%%%%%%%%%%%%%%%%%%%%%%%%%%%%%%%%%%%%%%%%%%%%%%%%%%%%%%%%%%%%%%%%
%  V.A   
%%%%%%%%%%%%%%%%%%%%%%%%%%%%%%%%%%%%%%%%%%%%%%%%%%%%%%%%%%%%%%%%%
\subsection{System Simulation}

\begin{figure}
  \includegraphics[trim=0 0 0 0,clip,width=\columnwidth]{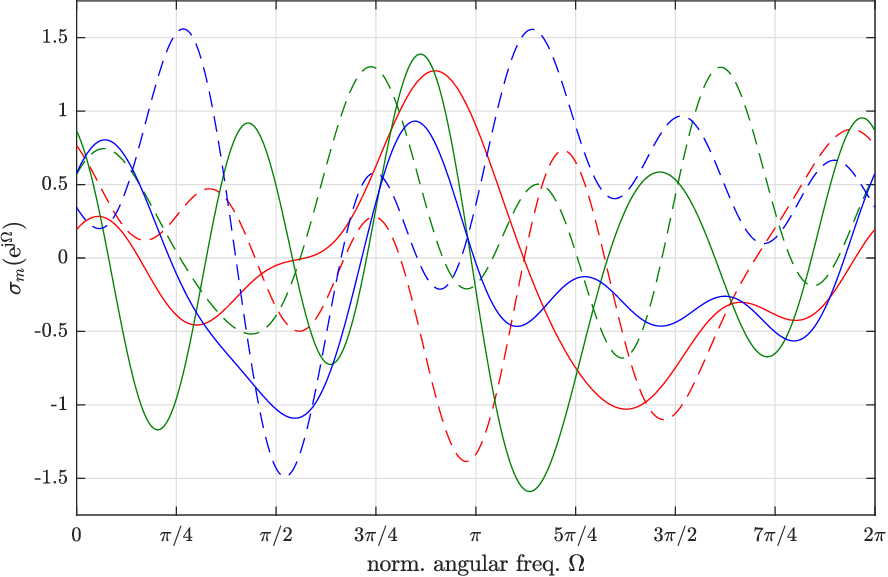}
  \caption{Ground truth singular values $\sigma_m(z)$, $m=1,\dotsc,6$ of
    system $\PMatrix{A}(z)$.
    \label{fig:BigSys}}
\end{figure}

In addition to the previous examples, we want to highlight the
perturbation of a more complicated matrix $\PMatrix{A}(z)$ by
estimation errors with different variance terms. For this, we
determine $\PMatrix{A}(z)$ from a given ground truth factorisation in
\eqref{svd_decomp}. The matrices holding the left- and right-singular
values are obtained form a concatenation of random elementary
paraunitary matrices~\cite{vaidyanathan93a}
\begin{align}
  \PMatrix{Q}(z) & = (\Matrix{I} - \Vektor{w}\Vektor{w}\Herm) +
  \Vektor{w}\Vektor{w}\Herm z^{-1} \; ,
   \label{eqn:Q_rand}
\end{align}  
where $\Vektor{w}$ is a random unit-norm vector such that
$\|\Vektor{w}\|_2 = 1$. With a concatenation of first-order terms drawn from random
instantiations of \eqref{eqn:Q_rand}, we create $\PMatrix{U}(z):
\Complex \rightarrow \Complex^{6\times 6}$ of order 10, and likewise
$\PMatrix{V}(z)$. The ground truth analytic singular values
$\sigma_m(z)$, which for real-valuedness of $\sigma_m(\ejo)$ must be
parahermitian, are created from random finite impulse responses
$s_m[n]$ of length 6 via
\begin{align}
  \sigma_m(z) = s_m(z) + s_m\PH(z) \; .
  \label{eqn:force_ph}
\end{align}
It is easy to check that with an arbitrary $s_m[n]$, with
\eqref{eqn:force_ph} we have indeed $\sigma_m(z) = \sigma_m\PH(z)$.
This generates a matrix $\PMatrix{A}(z): \Complex \rightarrow
\Complex^{6 \times 6}$ of order 26. For a particular instance, the
resulting singular values $\sigma_m(\ejo)$, $m=1,\dotsc,6$, are
illustrated in Fig.~\ref{fig:BigSys}.

%%%%%%%%%%%%%%%%%%%%%%%%%%%%%%%%%%%%%%%%%%%%%%%%%%%%%%%%%%%%%%%%%
%  V.B   
%%%%%%%%%%%%%%%%%%%%%%%%%%%%%%%%%%%%%%%%%%%%%%%%%%%%%%%%%%%%%%%%%
\subsection{Variable Perturbation}

\begin{figure}
  \includegraphics[trim=0 0 0 0,clip,width=\columnwidth]{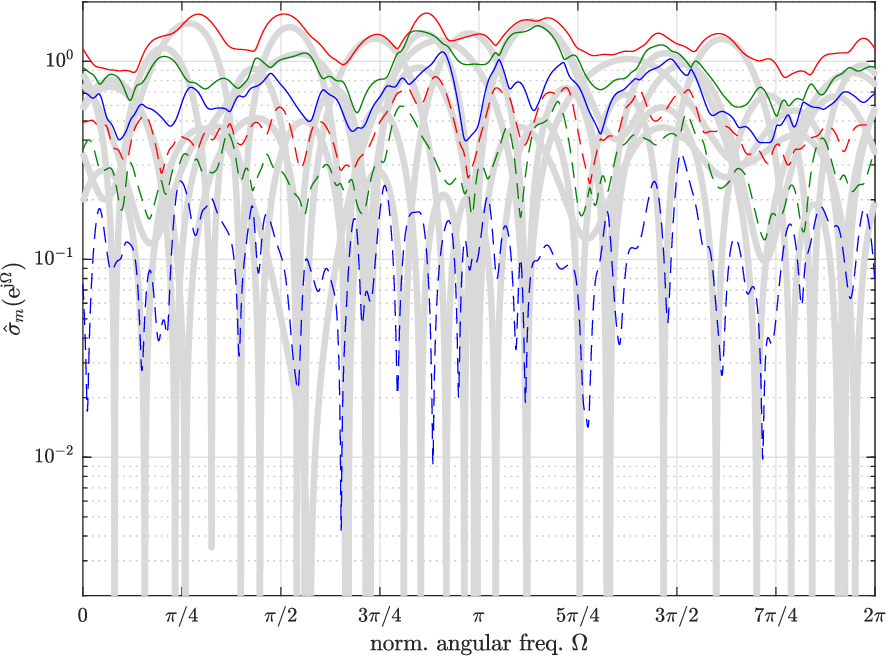}

  \vspace*{-.3cm}
  
  \caption{Analytic singular values $\hat{\sigma}_m(\ejo)$ of the
    perturbed system $\hat{\PMatrix{A}}(z)$, with
    $\sigma^2_{e,\mathrm{norm}} = 0.3$; the moduli of $\sigma_m(\ejo)$
    are underlaid in grey.
    \label{fig:BigSys2}}
  
    \vspace*{.3cm}

  \includegraphics[trim=0 0 0 0,clip,width=\columnwidth]{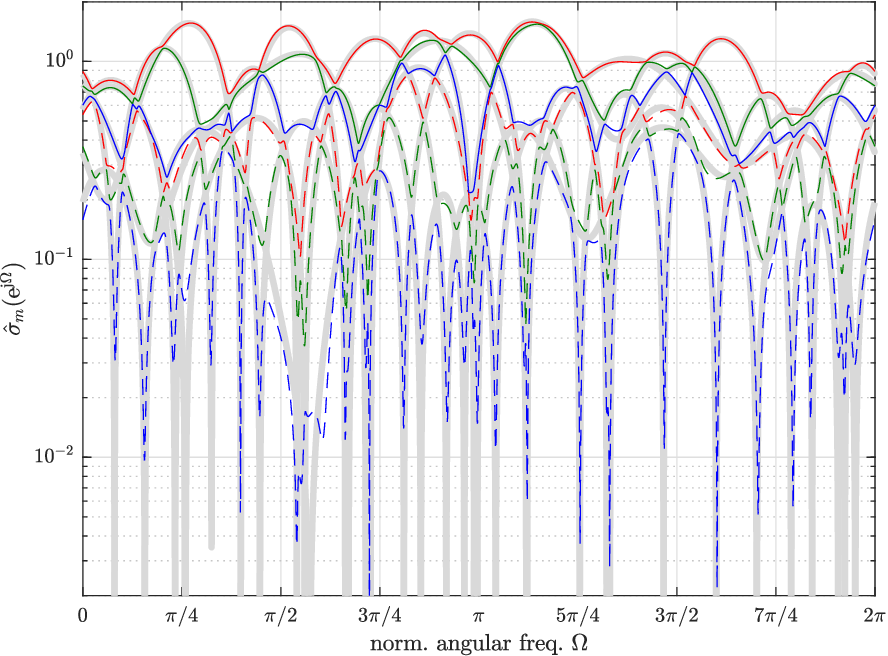}

    \vspace*{-.3cm}
  
\caption{Analytic singular values $\hat{\sigma}_m(\ejo)$ of the
    perturbed system $\hat{\PMatrix{A}}(z)$, with
    $\sigma^2_{e,\mathrm{norm}} = 0.01$; the moduli of $\sigma_m(\ejo)$
    are underlaid in grey.
    \label{fig:BigSys3}}

    \vspace*{.3cm}

\includegraphics[trim=0 0 0 0,clip,width=\columnwidth]{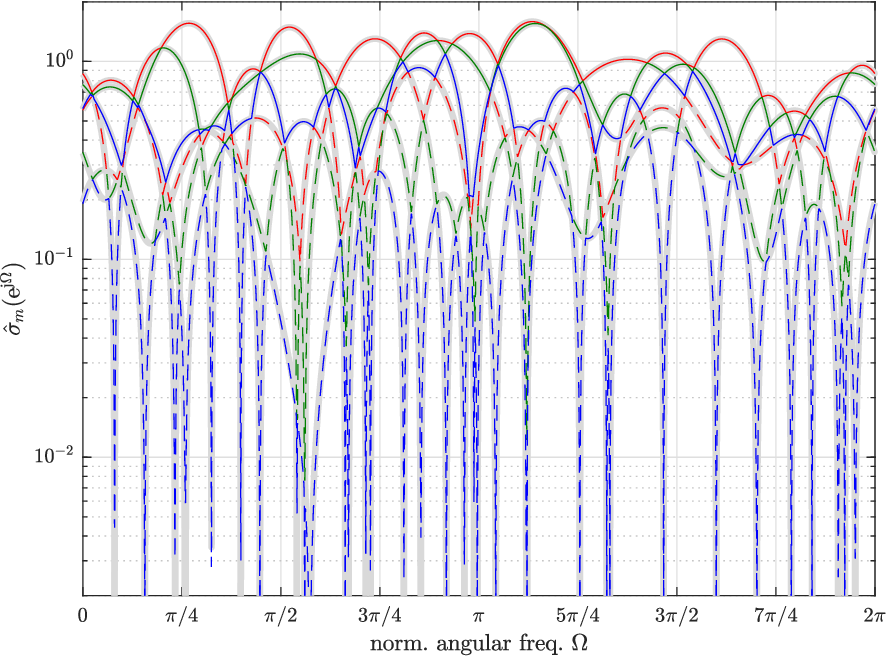}

  \vspace*{-.3cm}
  
\caption{Analytic singular values $\hat{\sigma}_m(\ejo)$ of the
    perturbed system $\hat{\PMatrix{A}}(z)$, with
    $\sigma^2_{e,\mathrm{norm}} = 10^{-4}$;  the moduli of $\sigma_m(\ejo)$
    are underlaid in grey.
    \label{fig:BigSys4}}
\end{figure}

We now simulate an estimated matrix $\hat{\PMatrix{A}}(z)$ by randomly
perturbing $\PMatrix{A}(z)$ by a term $\PMatrix{E}(z)$. The size and
order of $\PMatrix{E}(z)$ are selected to match the parameters of
$\PMatrix{A}(z)$, and similar to Example~\ref{ex:ex4}, its
coefficients are picked from a complex Gaussian distribution
$\mathcal{CN}(0,\sigma^2_e)$, but we will change the variance
$\sigma^2_e$, which is meant to represent the variance of the
estimation error. In order to relate this estimation error to the
energy in $\PMatrix{A}(z)$, we define the normalised variance
\begin{align}
  \sigma^2_{e,\mathrm{norm.}} = \frac{\sum_n \|\Matrix{E}[n]\|^2_{\mathrm{F}}}
    {\sum_n \|\Matrix{A}[n]\|^2_{\mathrm{F}}} \; .
\end{align}
For values of $\sigma^2_{e,\mathrm{norm}} = \{0.3; 10^{-2};
10^{-4}\}$, the singular values of the perturbed matrices
$\hat{\PMatrix{A}}(z) = \PMatrix{A}(z) + \PMatrix{E}(z)$ are provided
in Figs.~\ref{fig:BigSys2}, \ref{fig:BigSys3}, and \ref{fig:BigSys4}.

It is evident in Figs.~\ref{fig:BigSys2}--\ref{fig:BigSys4} that with
decreasing perturbation, i.e.~a smaller $\sigma^2_{e,\mathrm{norm}}$,
the singular values $\hat{\sigma}_m(\ejo)$ on a bin-wise view approach
the ground truth. They remain strictly spectrally majorised, i.e.~do
not intersect, and closer and closer approach zero where
$\sigma_m(\ejo)$ for $\sigma^2_{e,\mathrm{norm}} \rightarrow 0$. Yet,
the singular valyes will only reach and intersect zero in the limit for
$\sigma^2_{e,\mathrm{norm}}=0$.

%%%%%%%%%%%%%%%%%%%%%%%%%%%%%%%%%%%%%%%%%%%%%%%%%%%%%%%%%%%%%%%%%
%
%  VI. DISCUSSION AND CONCLUSIONS
%
%%%%%%%%%%%%%%%%%%%%%%%%%%%%%%%%%%%%%%%%%%%%%%%%%%%%%%%%%%%%%%%%%
\section{Summary and Impact
   \label{sec:concl}}
   
We have investigated how a random perturbation affects a matrix
$\PMatrix{A}(z)$, which may arise for example when estimating a matrix
of transfer functions, and how this perturbation impacts on its
analytic singular values. On the unit circle, where the analytic
singular values of $\PMatrix{A}(z)$ can have algebraic multiplicities
and zero-crossings, the analytic singular values
$\hat{\sigma}_m(\ejo)$, $m=1,\dotsc,M$, of $\hat{\PMatrix{A}}(z)$ will
have lost both the no-trivial algebraic multiplicities and the
zero-crossings with probability one. Thus, the analytic singular
values of the randomly perturbed system with be strictly spectrally
majorised and non-negative, even if the ground truth analytic singular
values are not. While for a decreasing perturbation, the bin-wise
singular values will converge to the ground truth, it is only in the
transition to a zero perturbation that the analytic singular values
will match the ground truth, and potentially be able to intersect and
change sign.

The loss of algebraic multiplicities and zero-crossings has a profound
impact on potential algorithms to extract the analytic SVD factors
from a perturbed matrix. If we pursue the extraction of
$\hat{\sigma}_m(z)$ directly, then this task will become the more
difficult the smaller the estimation error is. This is somewhat
counter-intuitive, as one would expect an estimate to be more useful
the more confident it is. However, since for a decreasing perturbation
the functions $\hat{\sigma}_m(\ejo)$ converge stronger towards
non-differentibilities where the ground truth $\sigma_m(\ejo)$
possesses intersections and zero-crossings, the approximation of
$\hat{\sigma}_m(z)$ generally requires higher and higher
polynomial orders for the SVD factors. Hence, direct approaches such
as for the analytic eigenvalue decomposition in~\cite{weiss21a,
  weiss23b} will become very costly, and a new methods akin
to~\cite{schlecht24a} need to be found that can target the less
complex ground truth rather than the exact SVD solution of the
perturbed system.

%%%%%%%%%%%%%%%%%%%%%%%%%%%%%%%%%%%%%%%%%%%%%%%%%%%%%%%%%%%%%%%%%
%
%  REFERENCES

\end{document}

\bibliographystyle{IEEEtran}
\footnotesize{
  \bibliography{
    %ES1,ES2,psvd_extra_refs,ES_extra_refs,
       /home/stephan/library/stephan_refAK_bib,
               /home/stephan/library/stephan_refLR_bib,
               /home/stephan/library/stephan_refSZ_bib}}


\end{document}